\documentclass[a4paper,12pt]{article}
\usepackage[T1]{fontenc}

\usepackage{psfrag}
\usepackage{graphicx}

\def\Journal#1#2#3#4{{#1} {\bf #2}, #3 (#4)}

\def\NPB{{\em Nucl. Phys.} B}
\def\PLB{{\em Phys. Lett.}  B}
\def\PRL{\em Phys. Rev. Lett.}
\def\PRD{{\em Phys. Rev.} D}
\def\ZPC{{\em Z. Phys.} C}

\begin{document}

\begin{titlepage}
\begin{center}
\vspace*{1cm}

\begin{title}
\bold
{\Huge Parameters in Weight Calculations \\for the BE Effect 

   }
\end{title}

\vspace{2cm}

\begin{author}
\Large
{K. Fia{\l}kowski\footnote{E-mail: uffialko@th.if.uj.edu.pl} and 
R. Wit\footnote{E-mail: wit@th.if.uj.edu.pl}
}
\end{author}

\vspace{1cm}
{\sl M. Smoluchowski Institute of Physics,

 Jagellonian University,

ul. Reymonta 4, 30-059 Krak\'ow, Poland}

\vspace{1cm}

\begin{abstract}
The weight method of implementing the BE effect into Monte Carlo generators
is discussed and presented in some detail. We show how the choice of free parameters
and the definition of "direct" pions influence the results for the hadronic $Z^0$ decays. 

\end{abstract}
\end{center}

\vspace{2cm}


\vspace{1cm}

\noindent

29 July, 2002 \\

\end{titlepage}

\section{Introductory remarks}
\par

   The effect of Bose-Einstein symmetrization (BE effect) in the
two-particle correlation spectra depends on the shape and size of the
source. This allowed to estimate the source parameters of
astronomical sources via the so-called Hanbury-Brown and Twiss effect
\cite{HBT}. The analogous  estimates in particle physics are much more
involved \cite{GGLP,BGJ,RMW}. In fact,  the applicability 
of the standard
analysis assuming incoherent production in particle collisions was
questioned recently and an alternative approach was presented \cite{BA}.  The
implementation of BE effect into Monte Carlo generators modelling
multiple production is particularly difficult, as the symmetrization
should be done at the level of amplitudes and generators deal
usually with probabilities. As far as we know there is only one
 implementation based on
the specific assumptions concerning amplitudes, and it applies only
for a single string fragmentation processes \cite{AR}.  The most widely
used procedure modelling the BE effect in the popular PYTHIA
generator \cite{PYT} is based on the prescription for shifting the final
state momenta to produce an enhancement at small momentum differences
in the distributions of pairs of identical hadrons \cite{LS}. In this procedure
one fits the parameters of the "input BE function"
\begin{equation}
F(Q) = 1 + \lambda exp(-Q^2R^2 )
\end{equation}
(assumed to have the same form as the standard parametrization
of BE effect) to reproduce the experimentally observed effect.
\par
There
is no simple relation between the values of input parameters $\lambda$
 and $R
$
and the analogous  parameters describing the experimental
distribution.  There is also no theoretical justification for this
procedure and it should be regarded as a convenient parametrization,
rather than the physical description of the BE effect.
\par  The alternative
approach is based on the formalism of Wigner functions \cite{Pratt}. One
approximates the corrected distribution as a product of distribution
without the BE effect and the weight function for which a definite
prescription is given \cite{BK}. This allows us to produce the distributions
with the BE effect by generating the events without this effect and
attaching to them the weights. To calculate these weights one must adopt
several simplifying assumptions \cite{BK} (and hope they do not destroy the
validity of the formulae). Finally one must assume the form of "two
particle weight factor" and fit its parameters to describe correctly
the data.
\par Superficially, there is a marked similarity between these two
approaches. In both cases the form of an "input function" is
assumed and its parameters should be fitted to describe the data.
However, there are also clear differences. Whereas the "momentum
shifting method" has no theoretical justification, it has two free
parameters (plus a few hidden parameters defining the choice of
momenta to be shifted, which affect the results rather mildly) and
is quite easy to apply. Since all generated events are used,
neither the multiplicity distributions nor, e.g., the decay channel
probabilities in $Z^0$ decay are affected by shifting. There seems to
be no need to change the values of the generator parameters fitted to
the data before taking the BE effect into account.  
\par On the other
hand, the weight method is quite well justified (granting that
simplifying assumptions are not too rough), but there are many
technical problems with its use. Some of them have been  solved:
prohibitive increase of computational time with multiplicity may
be avoided by a proper clustering procedure for final state
momenta \cite{FWW} and the distortion of the multiplicity distribution
may be removed by simple rescaling of weights depending on the
event multiplicity \cite{JZ}. Obviously, the weights may in principle
affect other distributions which were fitted to data without
taking the BE effect into account. Thus  the proper procedure would
be to refit all the generator parameters comparing the weighted results
with data. However, if the rescaling guarantees that
average weight is equal one for each well defined class of events
(e.g. in each $Z^0$ decay channel), the changes in distributions
should be minor. 
\par Another notorious technical problem for weight
methods is the instability of results due to the long tail of very
high weight values. Usually it requires some arbitrary cut, but
for sufficiently high number of generated events the effects of
this cut are not very significant.  Finally, there is a problem of
selecting the particles, whose momenta are used to calculate
weights and a problem of proper choice of "two particle weight
factor" and its parametrization (reflecting somehow the shape 
and the size
 of the production source).
\par 
In this
 paper we discuss the solutions to the last two
problems presenting the MC results for the BE effect in the hadronic decay of
$Z^0$ and comparing them with some data.
We consider only the distributions in the invariant four momentum difference 

\begin{equation}
Q^2 = - (p_1-p_2)^2.
\label{eq:Q2}
\end{equation} 

\noindent The effect of anisotropy in various components of $Q$ 
\cite{L3} was discussed elsewhere \cite{APP}.
\par 
 The following chapter contains the discussion of possible
particle selections and the influence of various MC parameters.
 The effects of different forms of two particle
weight functions (considered already in the earlier paper \cite{EPJC}) are
presented in the third chapter. The last chapter presents the
comparison with some data and conclusions.  It should be stressed
that we discuss only the standard prescription for weights justified
by the Wigner function formalism \cite{Pratt,BK}. Other proposals 
\cite{JZ,KKM,UWW} should
be regarded as viable versions of the weight method only if it is
shown that they reproduce approximately the results obtained for this
prescription.

\section{Particle selection and MC parameters}
\par
Before discussing the details of the MC procedure implementing the BE effect
we should decide which distribution will be used to present this effect.
The standard quantity called "the BE ratio" is defined as
\begin{equation}
R_{BE}(Q)=\frac{\rho_2(Q)}{\rho_2^0(Q)}
\end {equation}
where $Q^2$ was defined above (\ref{eq:Q2}) and the numerator and 
denominator represent the identical
two-particle distribution with and without the BE effect, respectively.
Obviously, this definition requires a more precise formulation of how we shall
define the denominator.
\par
 In the experimental definition of the BE ratio one uses often the distribution of unlike
sign pion pairs but this requires cutting off the resonance effects. Thus recently
 it is preferred
to use the pairs of identical pions from different events 
\begin {equation}
R_{BE}(Q)=C_2^{BE}(Q)\equiv \frac{\rho_2(Q)}{\rho_1 \otimes \rho_1 (Q)},
\label{eq:re}
\end{equation}
where the denominator is a convolution of single distributions 
\begin{equation}
\rho_1 \otimes \rho_1 (Q) = \int \rho_1(p_1)\rho_1(p_2)\delta (Q^2+(p_1-p_2)^2).
\end{equation}
This choice of the denominator has other flaws (i.e. it removes all
correlations, and not only the BE effect). Therefore  one
 uses often
 double ratios, dividing the experimental ratio by an
analogous ratio of distributions from MC generator (without the BE
effect)
\begin {equation}
R'_{BE}(Q)=\frac{C_2^{BE}(Q)}{C_2^{MC}(Q)}.
\label{eq:dre}
\end{equation}
For the MC generated events the simplest choice is just to run MC without the procedure
implementing the BE effect
\begin{equation}
R_{BE}^{MC}(Q)=\frac{\rho_2^{MC,BE}(Q)}{\rho_2^{MC}(Q)}
.
\label{eq:rmc}
\end {equation}
 Obviously if in the experimental investigation a double
ratio is used, it seems more proper to calculate for comparison an analogous  double ratio
from MC events

\begin {equation}
R_{BE}^{'MC}(Q)=\frac{C_2^{MC,BE}(Q)}{C_2^{MC}(Q)}.
\label{eq:drmc}
\end{equation}

Fortunately the difference between $R_{BE}^{'MC}(Q)$ and $R_{BE}^{MC}(Q)$ is often insignificant.
This is illustrated in Fig.1, where we show both ratios calculated for pion pairs from $Z^0$
decays using the weight method with the Gaussian two-particle weight factor

\begin{equation}
w_2(p_1,p_2)=exp\big(-\frac{(p_1-p_2)^2}{2\sigma}\big )
\label{eq:w2}
\end{equation}

\noindent 
with $\sigma = 0.05~GeV^2$.

\vspace{1.0cm}

\begin{center}

\includegraphics[width=11cm]{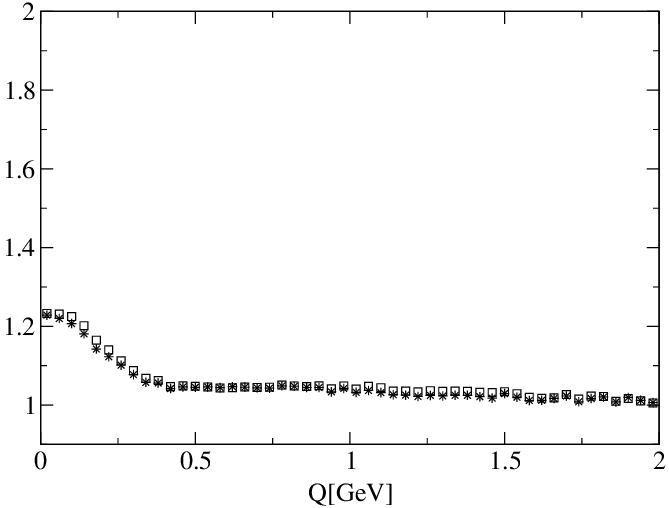}
\end{center}

\par {\bf Fig.1.} {\sl \small Comparison of the ratio of Q distributions (\ref{eq:rmc}) with and without
weights (squares) with double ratio of spectrum-to-background ratios
 (\ref{eq:drmc})
with and without weights (stars).}

\vspace{0.5cm}

Here, as in all the later figures:
\begin{itemize}
\item one million of events was generated
by the PYTHIA 6.2 generator \cite{PYT},
\item the background distributions were 
constructed using pairs from
different events; four million pairs of events were used for this purpose,

\item the BE ratios were normalized to approach smoothly the value of one at $Q$ exceeding $1~GeV$. 
\end{itemize} 
\par
For completeness, let us remind here that the two-particle weight factor
is related to the full weight calculated for each event by a formula
 \cite{BK}

\begin{equation}
W(p_1,...,p_n) =\sum \prod_{i=1}^n w_2(p_i,p_{P(i)})
\end{equation}

\noindent 
where the sum extends over all permutations of $n$ elements. More precisely,
the event weight is a product of such sums calculated for all kinds of identical
bosons registered by the detectors (in practice it is enough to count only positive
and negative pions).
\par
This formula suggests that the $\sigma$ parameter in formula (\ref{eq:w2})
is the only free parameter in the method. This would be, however, an oversimplification.
The calculation of the full sum over permutations is practically impossible for the number of
identical pions exceeding twenty \cite{HAY}. Thus we define the clusters of 
pions "close to each other"
in the phase space and sum over permutations within clusters only. To make the results
independent on the cluster definition we have to choose the value of the relevant
parameter $\epsilon$ much bigger than $\sigma$. The details of this procedure have been
described elsewhere \cite{FWW}.
\par
Moreover, if the two particle weight factor is interpreted as a Fourier transform
of the space--time distribution of pairs of pions, it seems justified to use a common
shape of this factor only for pairs of "direct" pions.

 The decay products of long
living particles and resonances are born far away from the original source and 
the corresponding
 two-particle
weight factor for pairs including these decay products would be close to the
 Dirac delta function, contributing negligibly
to the final event weight. 
The same reasoning was presented by Sj\"ostrand \cite{PYT}
who choose $20~ MeV$ as a limit  defining "long living" resonances and performed
the momentum shift only for pions produced directly and the decay products of
broader resonances.
\par
This suggests that we should calculate the event weight including in the sum only "direct"
pions defined in a similar way. To avoid the changes of the original Monte Carlo procedures
(which was the case for Sj\"ostrand PYBOEI procedure, called internally from the
generator before the decay of "long living" resonances and particles), we form a table
of momenta for "direct" pions defined in various ways and use this table for the weight calculation.
We found that the modifications of the original width limit of $20~MeV$ are irrelevant
as long as we do not include the $\omega$ decay products in the weight calculations.
Including $\omega$ decay products enhances strongly the BE ratio, as shown in Fig.2
for the $Z^0$ decay.
\par
To justify the choice of the width limit let us require that the "direct" pion and the pion from
$\omega$ decay have approximately the same momenta. The maximal momentum of a pion in the decay 
of $\omega$ at rest is about $300~MeV/c$,
and the most likely value is of the order of $100~MeV/c$.  This allows to estimate that the distance
between "birth points"
of such pions is of the order of $10~fm$ and the corresponding width in momentum space
should be about $20~MeV$,
smaller than the typical resolution. This suggests that the decay
products of $\omega$ (as well as the decay products of narrower resonances and other unstable particles)
should not be taken
 into account when calculating weights.

\vspace{1.0cm}

\begin{center}
\includegraphics[width=11cm]{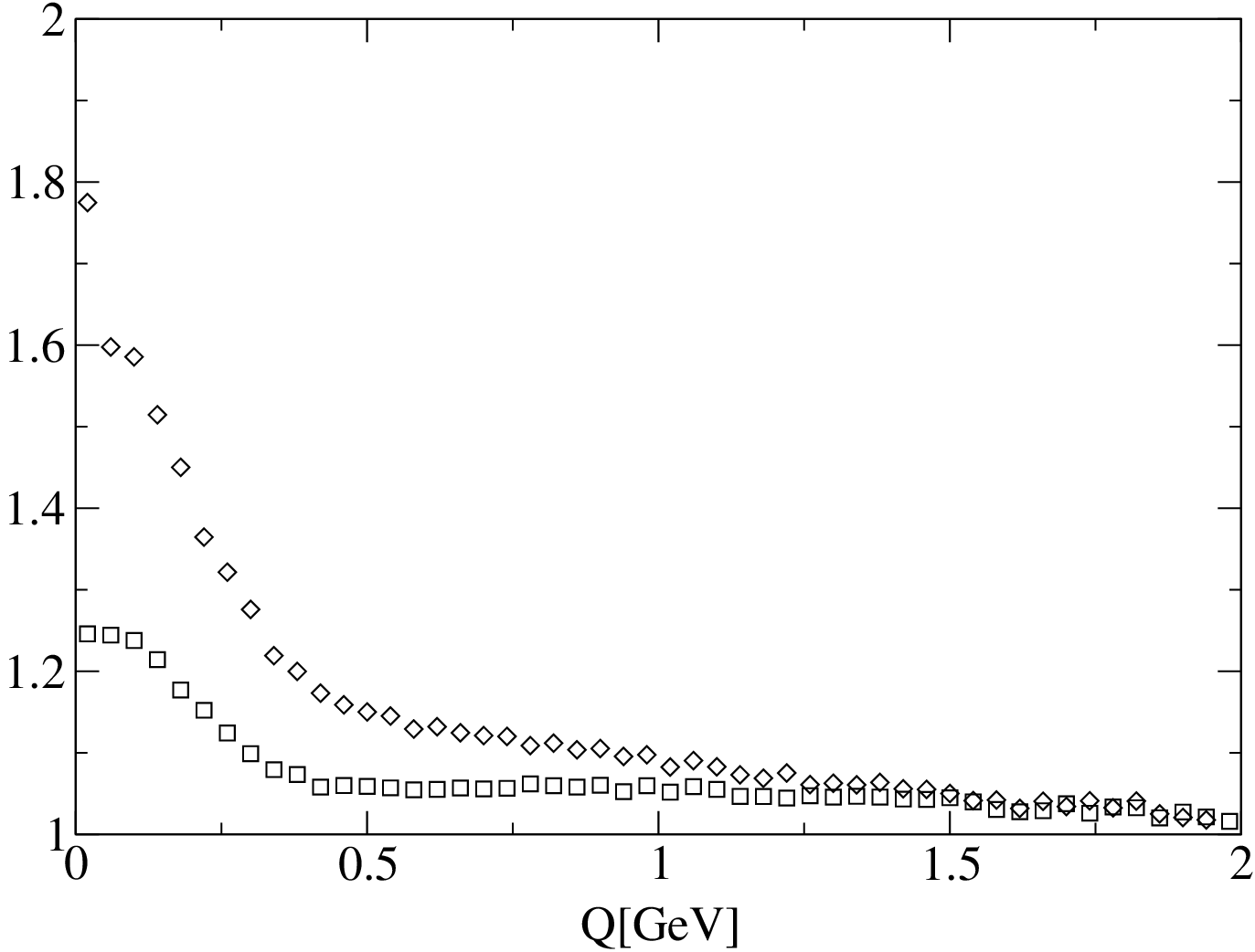}
\end{center}
\par {\bf Fig.2.} {\sl \small Comparison of the ratio of Q distributions  with and without
weights (\ref{eq:rmc}) for weights calculated excluding (squares) and including (diamonds)
$\omega$ decay products.}

\vspace{0.5cm}

\par
However, this argument has some flaws. First, the BE ratio is defined as a
function of $Q^2$ and not of the
three-momentum squared (thus it reflects the space-time and not just the space
structure of source). Second, the distribution of the decay length is not Gaussian.
Thus we should not expect a Gaussian 
shape of the weight factor. Finally, excluding the decay products of narrow
resonances is a very
rough procedure; a better solution would be to use different two-particle weight
factors for different
pairs of pions (direct--direct, direct--resonance and resonance--resonance). 
Let us add that all this estimate is classical and does not take into
account
the possible quantum effects. Thus we should not treat the choice of "direct"
pions excluding the $\omega$
decay products as definite. In fact, the uncertainty of this choice seems to be
the biggest uncertainty
of the weight method. If necessary, it may be used to describe the BE effect 
if the observed values of the BE ratio at
 small $Q^2$ are high.

\par
The other free parameters of the PYTHIA generator may also influence the weights and the resulting BE ratio.
An example of this effect is shown in Fig.3 where we compare the results for default values of PYTHIA parameters and for the
values fitted to the L3 data \cite{L3}. Let us stress that the choice of "direct" pions (excluding the $\omega$
decay products) and the value of $\sigma$ parameter are the same in both cases, but the results are visibly
different. This is probably mainly due to the suppression of $\eta$ and $\eta'$ mesons for the L3 parameters.

\vspace{1.0cm}

\begin{center}

\includegraphics[width=11cm]{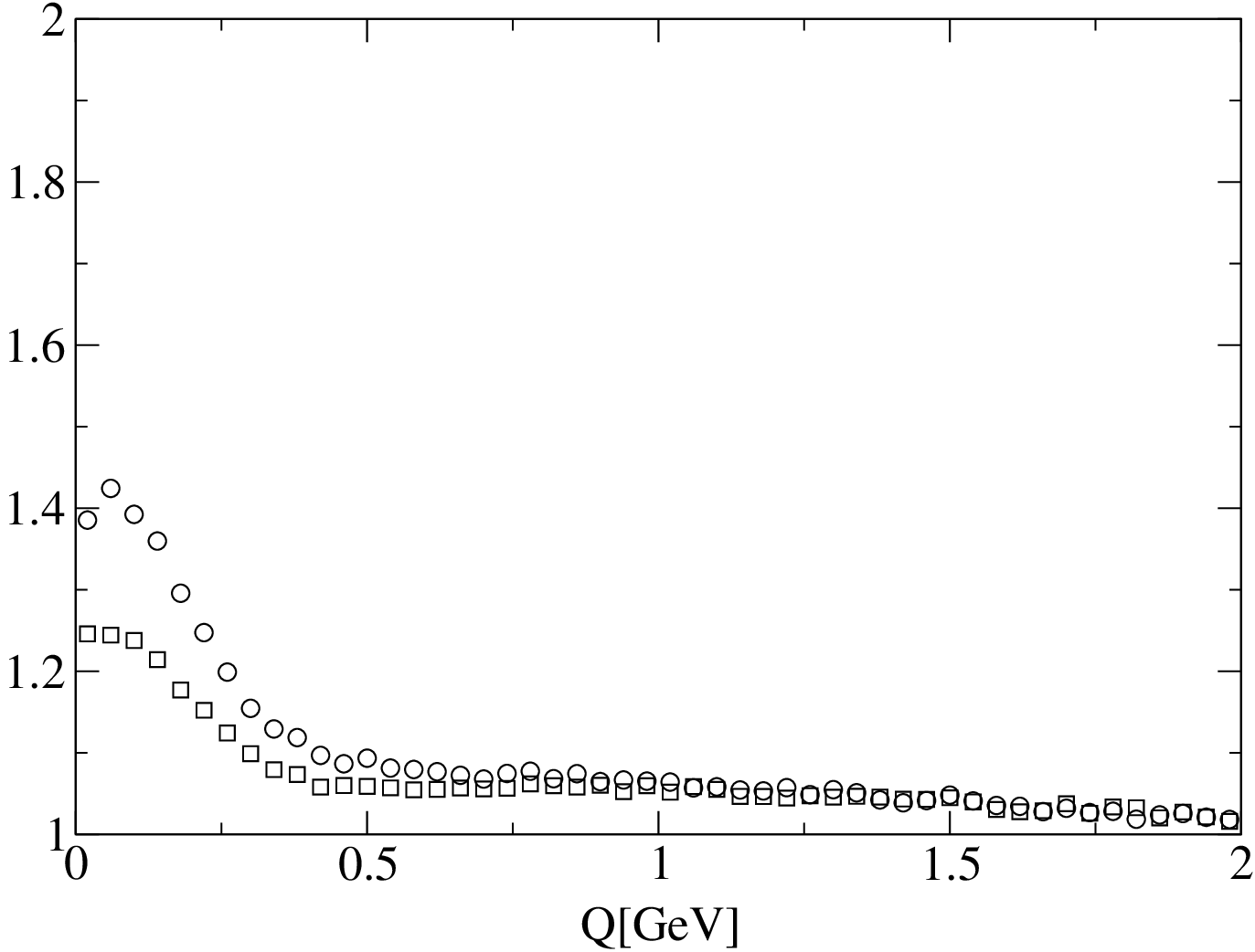}
\end{center}
\par {\bf Fig.3.} {\sl \small Comparison of the ratio of Q distributions  with and without
weights (\ref{eq:rmc}) for weights calculated using default PYTHIA parameters (squares)
and L3 parameters (circles).}
\vspace{0.5cm}
\par
Finally, let us check the dependence of the results on the value of the $\sigma$ parameter. Until now
we were using the value of $0.05~GeV^2$ which corresponds to the average Gaussian source size of
the order of $1~fm$. In Fig.4 we compare the results (with L3 parameters) for $\sigma =0.05~GeV^2$ and
$\sigma =0.07~GeV^2$. We see that by increasing $\sigma$ (which corresponds to a decreasing source size)
we increase the width of "BE peak" and slightly increase its height.

\vspace{1.0cm}

\begin{center}

\includegraphics[width=11cm]{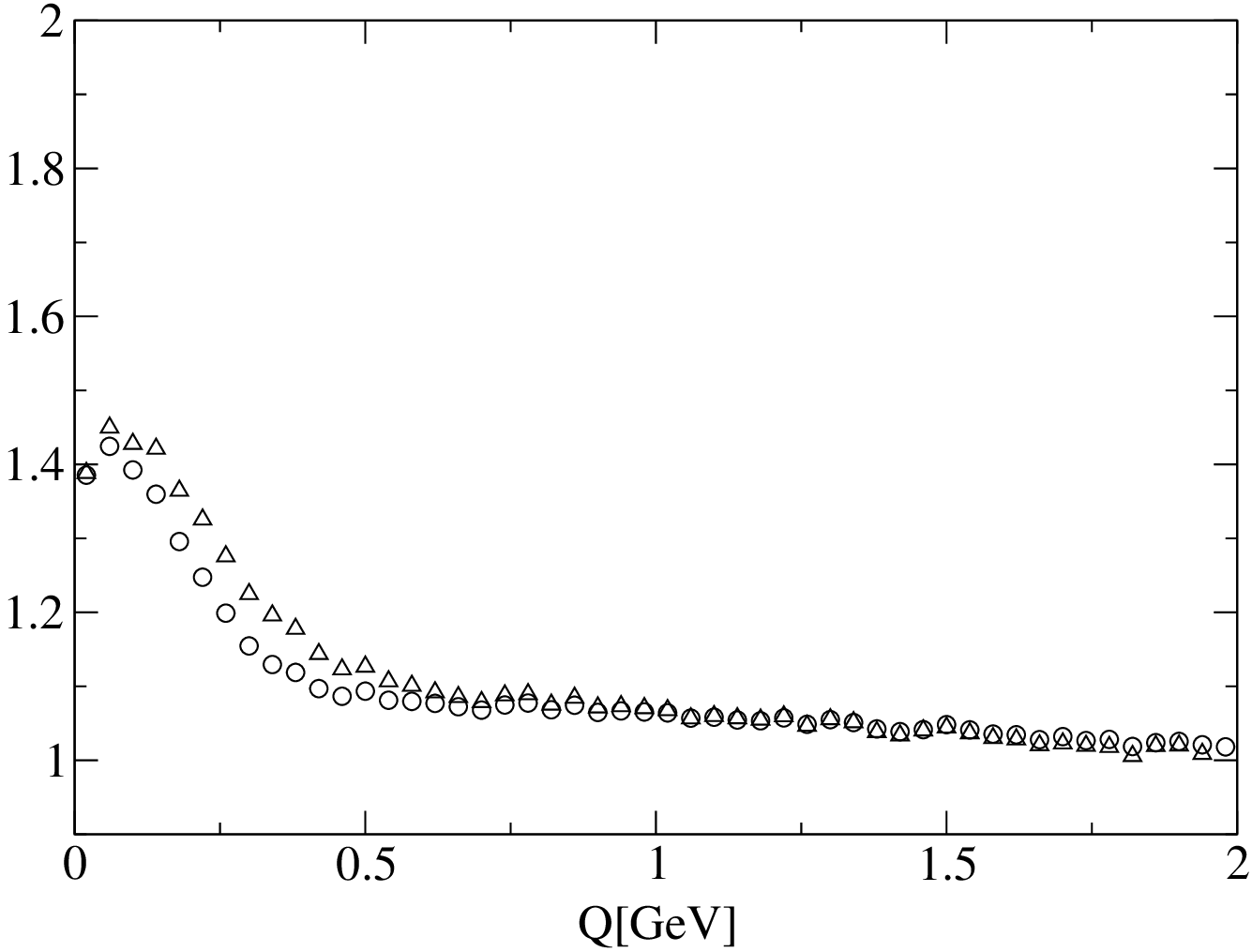}
\end{center}
\par {\bf Fig.4.} {\sl \small Comparison of the ratio of Q distributions  with and without
weights (\ref{eq:rmc}) for L3 parameters with weights calculated using $\sigma=0.05~GeV^2$
(circles) and $\sigma=0.07~GeV^2$ (triangles).}

\vspace{0.5cm}
\par
A notorious problem of the weight method is a possible distortion of various distributions
(fitted previously to data) by the introduction of  weights. First such a distortion 
was observed for the multiplicity distribution where the probabilities of 
high multiplicities were enhanced 
by weights. This was cured by rescaling the weights with a factor $C\Lambda^n$ \cite{JZ} where $n$ 
is a charge particle multiplicity (measured in experiment). The values of parameters 
$C$ and $\Lambda$ are fitted to restore the original values of $\overline n$ and the original 
normalization. With this method the weights cause only a moderate increase of the 
dispersion of the multiplicity distribution. 
\par 
The weights influence also the single particle momentum distribution, reducing slightly the width,
but these effects are not very significant. More important is the change in the $Q^2$ 
distribution of unlike sign pairs of pions, as shown in Fig.5. A similar effect 
was observed for  Sj\"ostrand's  implementation method of the BE effect. It should be noted, however,
that by including the $\omega$ decay products in the weight calculation we increase the $R$ ratio for
unlike sign pion pairs by a few percent only, whereas the ratio for like sign pairs increased by
about $50\%$, as shown in Fig.2. Thus it is possible to describe a big BE effect without distorting
seriously the distribution for unlike sign pairs.

\vspace{1.0cm}

\begin{center}

\includegraphics[width=11cm]{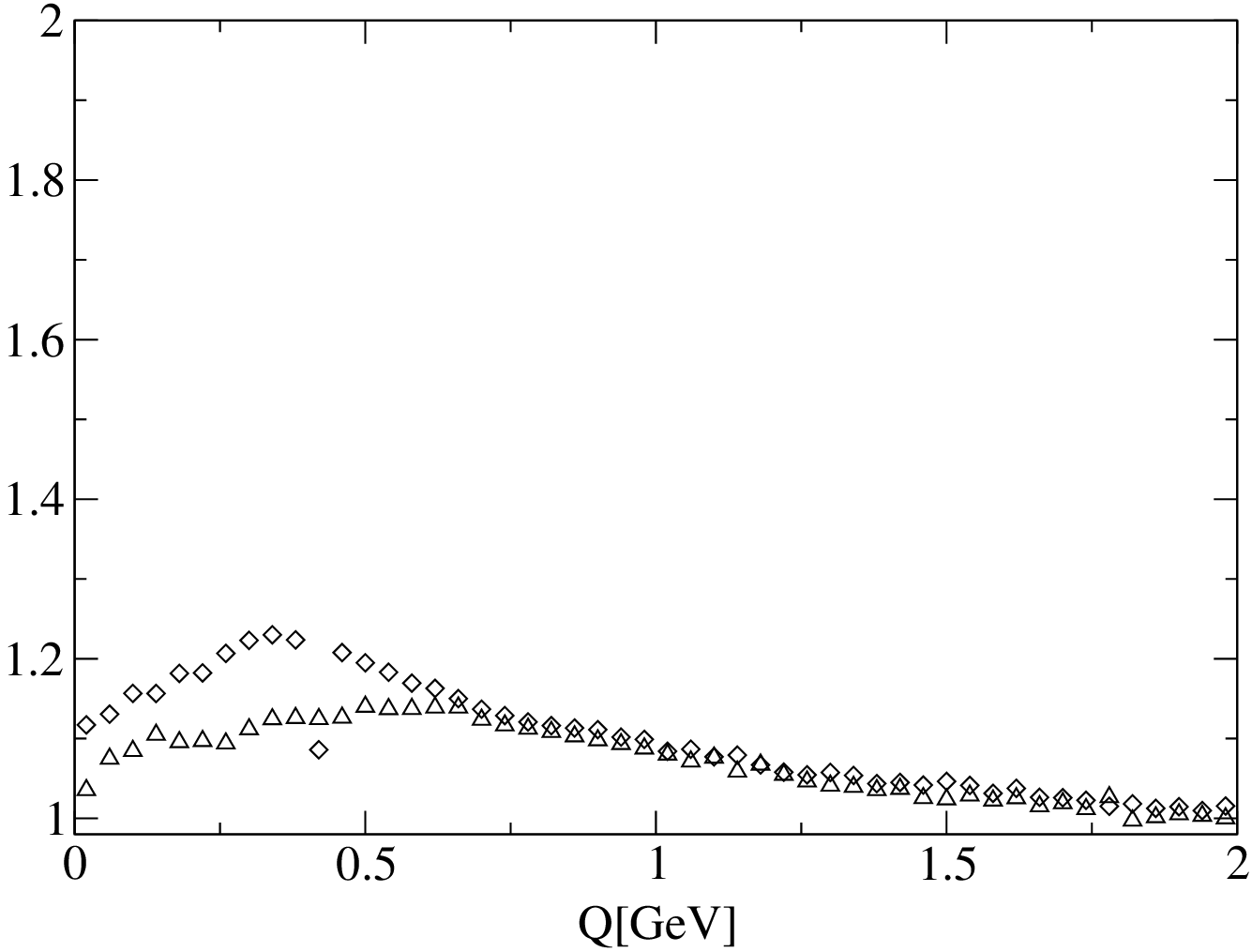}
\end{center}
\par {\bf Fig.5.} {\sl \small Comparison of the ratio of Q distributions (\ref{eq:rmc}) with and without
weights for unlike sign pairs of pions. Weights are calculated using L3
parameters with $\sigma=0.07~GeV^2$ excluding $\omega$ decay products (triangles)
and using default parameters with $\sigma=0.05~GeV^2$ including $\omega$ decay
products (diamonds).}
\vspace{1.0cm}

\section{Choice of the two-particle weight factor}
In the former section we used always the Gaussian two-particle weight factor (\ref{eq:w2}).  Obviously,
there is no reason why all the space-time and momentum distributions should be described by such
simple functions. However, if we restrict ourselves to the monotonically decreasing weight factors 
normalized to one at $Q^2=0$, it is easy to show that the Gaussian choice results in a curve which 
is the smoothest one and resembles the data best. This is demonstrated in Fig.6 where 
we compare the results
obtained for the default PYTHIA parameters for the Gaussian weight factor (\ref{eq:w2})
with $\sigma = 0.05~GeV^2$ and for two other choices of the weight factor: 
\begin{itemize}
\item step-like 
\begin{equation}
w_2(p_1,p_2)=\Theta[(p_1-p_2)^2+\sigma]
\label{eq:step}
\end{equation}
with the same value of $\sigma$
\item exponential
\begin{equation}
w_2(p_1,p_2)=exp\Big(-\frac{\sqrt{-(p_1-p_2)^2}}{2\sqrt{\sigma}}\Big)
\label{eq:exp}
\end{equation}
with $\sigma=0.03~GeV^2$.
\end {itemize}

\vspace{1.0cm}

\begin{center}

\includegraphics[width=11cm]{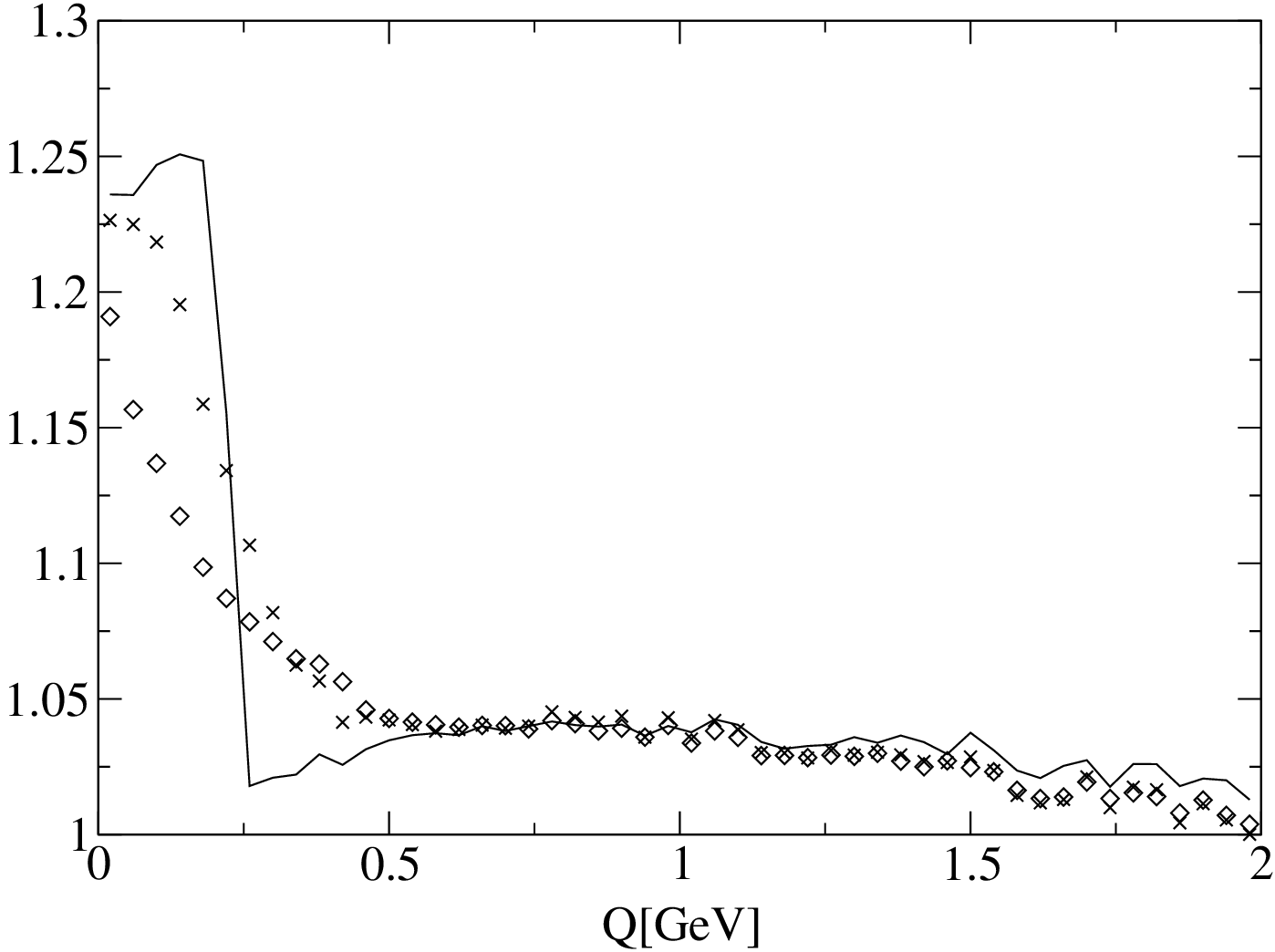}
\end{center}

\par {\bf Fig.6.} {\sl \small Comparison of the ratio of Q distributions (\ref{eq:rmc}) for default PYTHIA 
parameters and three different choices of two-particle weight factor: gaussian (crosses),
step-like (solid line) and exponential (diamonds).}

\vspace{0.5cm}

\par
We see clearly that the shape of the obtained  BE ratio reflects the shape of the two-particle 
weight factor. This may be written as an approximate relation 
\begin{equation}
R_{BE} (-(p_1-p_2)^2) \approx 1 + c\cdot w_2(p_1,p_2)
\label{eq:approx}
\end{equation}
where the value of $c$ depends on the shape of the weight factor and the 
selection of particles used in the weight calculations. 
\par
Let us note that the Gaussian parametrization is unlikely to describe 
the data where the distribution of the space-time   distance between the
"birth points" of two pions is more complicated. This is the case for the 
four jet decay of $W^+W^-$ final states if two pions originate from two 
$W$-s. There it is unjustified to expect monotonically decreasing and
normalized two-particle weight factors.
However, for the $Z^0$ decay the parametrization (\ref{eq:w2})
 seems to be the appropriate
one. 
 
\section{Comparison with data and conclusions}
We will not attempt here a detailed fit to any published data. There are many reasons 
for this reservation. First, as we have already mentioned, different experiments use
different definitions of the reference sample in the denominator of the BE ratio.
Thus the fit quality may depend on many factors not related to the procedure implementing 
the BE effect (e.g., the resonance effects and other correlations). Second, the published data 
include usually the acceptance corrections which are difficult to reconstruct in our calculations. In
fact, the Monte Carlo parameters should be also fitted to the particular set of data 
before implementing the BE effect. As shown in Section 2 there is a difference between 
the results obtained using default PYTHIA parameters and the parameters used by 
the L3 collaboration.

\par
Therefore we want to make only a semi-quantitative comparison between the results  from our procedure
and some high statistics data. To this purpose we use the recent L3 data shown as the reference 
sample in the paper devoted 
to the analysis of the $WW$ decay \cite{L32}. We compared them with various MC results shown in previous 
sections, rescaled with arbitrary constants to agree with data at $Q^2>1 GeV^2$. We found
that the modelled BE effect is too small compared with the data unless we include 
the $\omega$ decay products for the weight calculation.
We show the comparison in Fig.7 for two choices of the $\sigma$ parameter in the weight factor 
(\ref{eq:w2}). Normalization of both curves was adjusted to fit the data.
We see that the data are qualitatively described by the PYTHIA  MC with our 
implementation of the BE effect. One should not expect a good quantitative fit to the data
for any single value of $\sigma$; as already noted, one could use at least different values
of this parameter (and, even better, different shapes of $w_2$) for  pairs of 
pions of different origin. Then, however, the number of free parameters would increase 
making the success of the fitting procedure rather trivial.
\vspace{1.0cm}

\begin{center}

\includegraphics[width=11cm]{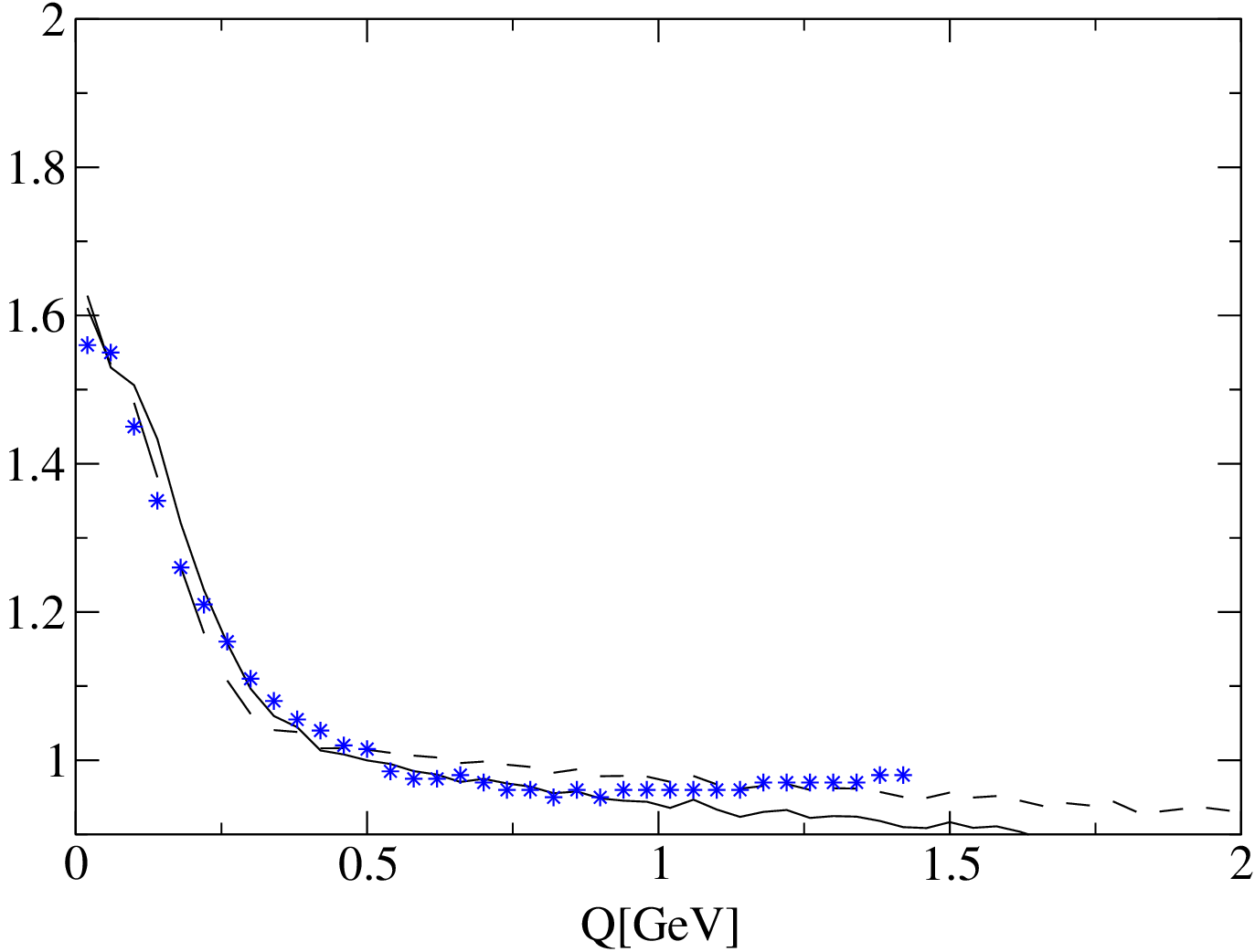}
\end{center}

\par {\bf Fig.7.} {\sl \small Comparison of the ratio of Q distributions for L3 parameters (\ref{eq:rmc})
with $\sigma=0.04$ (solid line) and $\sigma=0.03$ (broken line) with the L3 data (stars).}

\vspace{0.5cm}
To summarize, we have discussed the freedom of the weight method for implementing  
the BE effect into Monte Carlo generators. We have shown that this freedom seems to 
be sufficient to describe the data. For pions coming from a single source which may 
be parametrized with a Gaussian distribution, there are three steps for choosing the
weight method parameters:
\begin{enumerate}
\item One should decide which ratio is used to display the BE effect and to calculate 
the same ratio from the MC with weights. It is preferred to use double ratios (cf. (\ref{eq:dre}) 
and (\ref{eq:drmc})) where both for data and MC one divides the chosen BE ratio by 
the same ratio calculated from MC without weights.
\item One should choose the selection criterion for pions used to calculate weights.
The typical choice corresponds  to using direct pions and decay products 
of broad resonances, $\Gamma > 20~MeV$ (as in Sj\"ostrand's method). 
\item One should select a proper value of the parameter $\sigma$ in (\ref{eq:w2}).
\end{enumerate}
\par
Technically, our algorithm contains four Fortran procedures:
\begin{itemize}
\item LWBOEI, where for each event the "direct" pions are selected and their momenta are stored in the tables,
\item KLASKF, where pions of one sign are assigned to clusters,
\item PERCJE, where a weight factor from each cluster is calculated,
\item CLUSWAGI, where the full event weight is calculated as a product of weight factors from all clusters and
all pion signs.
\end{itemize}
All these procedures are available at request from us, together with a sample program calling the
PYTHIA 6.2 generator and comparing the weighted and unweighted distributions for hadronic $Z^0$ decays.
A modification of this program for other processes or other MC generators would be straightforward.
\par 
One should also remember that after the introduction of weights one should rescale them by 
a $C\Lambda^n$ factor to restore the original normalization and average multiplicity.
This, however, does not influence significantly the shapes of the BE ratios.
\par
To describe the process in which pions originate from two or more independent sources
(as the $e^+e^-\rightarrow W^+W^-$ process with double hadronic decay of $W$-s) one 
needs a more elaborate procedure. Different forms of the $w_2$ factor should be used for pairs 
coming from the same and from different sources. This will be discussed in detail elsewhere.

\section*{Acknowledgments}

We are grateful to A. Kota\'nski for reading the manuscript. 
This work was 
partially supported by the KBN grants 2P03B 093 22  and  
2 P03B 019 17  in 2002.

\end{document}